# High on-off Conductance Switching Ratio in Optically-Driven Self-Assembled Conjugated Molecular Systems

*Kacem Smaali (1), Stéphane Lenfant (1)\*, Sandrine Karpe (2), Maïténa Oçafrain (2), Philippe Blanchard (2), Dominique Deresmes (1), Sylvie Godey (1), Alain Rochefort (3), Jean Roncali (2) and Dominique Vuillaume (1)\**

(1) Molecular Nanostructures & Devices Group, Institute for Electronics Microelectronics and Nanotechnology, CNRS & University of Lille, B.P. 60069, 59652, Villeneuve d'Ascq, France.

(2) Linear Conjugated Systems Group, CNRS, CIMA, University of Angers, 2 Bd Lavoisier, 49045, Angers, France.

(3) Département de génie physique, École Polytechnique de Montréal, Montréal, Canada H3C 3A.

\* Send correspondence to dominique.vuillaume@iemn.univ-lille1.fr; stephane.lenfant@iemn.univ-lille1.fr

**Abstract.** A new azobenzene-thiophene molecular switch is designed, synthesized and used to form self-assembled monolayers (SAM) on gold. An "on/off" conductance ratio up to $7\times10^3$ (with an average value of $1.5\times10^3$) is reported. The "on" conductance state is clearly identified to the *cis* isomer of the azobenzene moiety. The high "on/off" ratio is explained in terms of photo-induced, configuration-related, changes in the electrode-molecule interface energetics (changes in the energy position of the molecular orbitals with respect to the Fermi energy of electrodes) in addition to changes in the tunnel barrier length (length of the molecules). First principles DFT calculations demonstrate a better delocalization of the frontier orbitals, as well as a stronger electronic coupling between the azobenzene moiety and the electrode for the *cis* configuration over the *trans* one. Measured photoionization cross-sections for the molecules in the SAM are close to the known values for azobenzene derivatives in solution.

The immobilization of π-conjugated systems with electro-optical properties on gold surfaces by formation of self-assembled monolayers (SAMs) is the focus of a high interest.[1] Reversible photoswitching devices were demonstrated with diarylethene and azobenzene derivatives,[2] and open the route for potential applications as optical switches in molecular electronics,[3-7] molecular machines,[8] photo-switchable surface wettability[9, 10] and biosensors.[11] Azobenzene molecules were extensively studied for their unique photoisomerization behavior. These molecules show a transition from a more thermodynamically stable *trans* configuration to a *cis* configuration upon exposure to UV light (~ 360 nm), and a reversible isomerization under blue light (~ 480 nm). The properties of azobenzene in solution (*e.g.* robust reversible photoisomerization, long living states, fast switching) make them promising building blocks for molecular-scale devices and nanotechnology.

It is well known that photoisomerisable molecules need to be electronically decoupled from the metal surface to properly work, *i.e.* to reversibly switch between the two isomers. STM experiments on a single azobenzene molecule physisorbed on gold show that reversible switching is only observed when tert-butyl legs lift the molecule up.[12] In several works using SAMs, the photoisomerisable molecules are chemically attached to the substrate using various spacers: short alkyl chains (2 to 6 carbon atoms),[3, 13, 14] ethylene bond[7] or phenyl or thiophene moieties.[2, 4, 8, 15, 16] The role of this linker is crucial. In the case of diarylethene, the observation of reversible switching depends on its nature (*e.g.* phenyl vs. thiophene)[16]. Contradictory, some authors associate the higher "on" conductance state to the *trans* isomer,[3, 7, 14, 17-20] while others conclude in favour of the *cis* form.[4, 8, 12] In some cases, this discrepancy can be accounted for by a different geometrical orientation of the molecules (lying flat on the substrate[12] or almost normal oriented), but the contradiction still holds for the same orientation (*e.g.* approximately standing upright normal to the electrode surface), see for instance refs. [17, 21]. Moreover, up to now, azobenzene derivatives do not exhibit a clear intrinsic conductance switching. The apparent change in the measured conductance has been attributed to a change in the length of the molecule during the isomerization rather than to an intrinsic conductance switching associated with changes in the electronic structure of the molecular junction.[4, 8] Many reasons can explain these results; details on the molecular arrangements in the monolayers and the nature of the coupling between the molecule and the electrode contact are among the most important factors that can influence the electrical behavior. For instance, Cuniberti *et al*.[17, 21] showed by first principles calculations that $G_{trans} > G_{cis}$ when the azobenzene is chemically linked between two carbon nanotube electrodes, while $G_{cis} > G_{trans}$ in case of silicon electrodes ($G_{trans}$ and $G_{cis}$ are the conductances of the junction for the *trans* and *cis* isomers, respectively). Obviously, the role of the spacer is also critical. A short spacer should favour the electron transfer rate through the junction and increases its conductance, while a longer spacer could improve the decoupling of the azobenzene moiety from the substrate, thus allowing a larger dynamic of the switching event, and thus a larger "on/off" conductance ratio.

Here, we report the electrical properties of a new molecular switch in which the azobenzene moiety is linked to a bithiophene spacer and a short (4 carbon atoms) alkanethiol. Such a design is expected to combine the benefit of a rather long spacer, while preserving a sufficiently high level of current due to the presence of electron-rich bithiophene unit (compared to a fully saturated spacer with the same length). The effects of the reversible photo-isomerization on the SAM thickness, surface wettability, and electrical conductivity are analyzed. The photo-induced changes in the conductance of the SAM are mainly investigated by conducting AFM (C-AFM). A record on/off ratio up to $7 \times 10^3$ between the *cis* ("on") and *trans* ("off") configurations is demonstrated. The analysis of these results using well-established electron transport models and molecular frontier orbitals from first principles DFT calculations indicates that this high photo-induced on/off ratio results from a synergistic combination of SAM thickness variation and modification of the energy offset between the lowest unoccupied molecular orbital (LUMO) and the electrode Fermi energy. We also report photoionization cross-sections for these molecules embedded in the molecular junctions on a par with known values for azobenzene and derivatives in solution.

**Monolayer formation and characterization.**

The new azobenzene-bithiophene derivative (AzBT) was synthesized and used to form self-assembled monolayers (SAM) by chemisorption on gold surfaces (see experimental section) - Figure 1. The SAMs were structurally characterized and all these measurements converge to identify the presence of dense monolayers of the desired molecules ($4\times10^{-10}$ mol/cm$^2$ or $2.4\times10^{14}$ molecules/cm$^2$ or 1 molecule *per* 41 Å$^2$ as determined by cyclic-voltammetry (CyV) measurements, see experimental section). The entire X-ray photoemission spectroscopy (XPS) spectrum (not shown here) of the functionalized surface shows the presence of the different atoms of the molecule: carbon (peak $C_{1s}$), nitrogen (peak $N_{1s}$), and sulphur (peaks $S_{2s}$ and $S_{2p}$). The $C_{1s}$ spectrum shows one component attributed to C–C at 284.3eV. This peak position is in good agreement with results published for monolayers grafted on gold where the position of the C–C peak has been measured at 284.6 eV.[22-24] The other peak positions ($N_{1s}$ and $S_{2s}$) appear at 399.5eV and 227.4eV, respectively. The ratio of the corrected peak area $N_{1s}$/$C_{1s}$ was about 0.082, a value close to the expected value 2/26=0.077 (2 nitrogen atoms and 26 carbon atoms per molecule). The high resolution XPS spectra for the $S_{2p}$ region shows two doublets (Figure 2). Each doublet results from the spin–orbit splitting of the $S_{2p}$ level and consists of a high-intensity $S_{2p3/2}$ peak at lower energy and a low-intensity $S_{2p1/2}$ peak at higher energy separated by 1.2eV with an intensity ratio of 2/1.[22] The value of the $S_{2p3/2}$ peak at 161.5 eV for the SAM is in agreement with the binding energy for bound alkanethiolate on gold (161.9–162.0 eV).[25, 26] Thus, peaks at 161.5 and 162.7 eV of the lower-energy doublet $S_{2p3/2,1/2}$ of SAM may be assigned to the thiol chemisorbed on the Au surface.[22, 27] The other doublet, $S_{2p3/2,1/2}$, at respectively 163.1 and 164.3 eV corresponds to other sulfur atoms in the body of the molecule (thiophene). The comparison of the areas of these two doublet signals leads to an estimated S-C/S-Au ratio of 6.2 ± 2.0 which is compatible with the expected value (S-C/S-Au = 4) for 1 sulphur atom linked to the gold surface versus 4 sulfur atoms linked with carbon atoms in the molecule.

We focalized UV-light (360 nm) and blue light (480 nm) on the SAM from a xenon lamp through an optical fiber (see experimental section). We followed the evolution of the water contact angle of the SAM as a function of the light irradiation wavelength (figure 3). Before light irradiation, the contact angle on the SAM is 94 ± 2°. After UV irradiation (*trans-to-cis* isomerization) during 90 minutes, the contact angle decreases to ~ 92 ± 2°, and switches back to a higher value (~ 98 ± 2°) after blue irradiation (*cis-to-trans* isomerization). This behavior is reproducible. The intermediate value for the pristine SAM can be explained by the fact that the SAM after its formation is composed of a mix of *cis* and *trans* isomers. The difference of wettability between the two isomers is due to several reasons. i) In the *trans* form, the surface of the SAM is mainly composed of $CH_3$ groups, while with the molecules in the *cis* form, the surface exposes mainly the N=N bonds and the phenyl groups. ii) Difference in the molecule dipole: the dipole moment of the *cis* isomer is higher than the *trans* isomer (for our molecule $\mu_{cis}$=4.9D and $\mu_{trans}$=1.8D from MOPAC calculations).[28] This variation of the contact angle between the two isomers, $\Delta\theta = 6 \pm 2°$, is close to those measured on other azobenzene systems.[9, 29]

Similarly, Fig. 3b shows the variations of the thickness for the SAM with molecules in the *cis* and *trans* isomers. The average thickness of the SAM after blue light exposure (*trans* isomer) is $d_{trans}$ = 30 ± 1 Å and $d_{cis}$ = 25 ± 1 Å after UV-light exposure (*cis* isomer). We compared these thicknesses with the lengths of the molecule calculated by MOPAC.[28] The length for the molecule in *cis* and *trans* forms is estimated ~ 26 and 31 Å, respectively. These values are very close to measured ones. The thickness of the pristine SAM is intermediate between the thickness of the SAM with molecules in their *cis* and *trans* conformations. This feature is consistent with a SAM composed of a mix of *cis* and *trans* isomers, as observed for the contact angle measurement above. The difference of thickness (here ~ 5 Å) is consistent with other results on azobenzene monolayers.[4, 6, 8, 9, 30] We can crudely estimate the ratio of *cis*/*trans* isomers in the pristine SAM by considering that the measured thickness is proportional to the relative amount of both isomers ($d = \eta d_{cis} + (1-\eta) d_{trans}$) as previously done in binary SAMs of alkylchains with two different lengths.[31] We get ~ 55% of *cis* isomers (with a relative accuracy of 25% corresponding to the ± 1 Å error bar of the ellipsometer measurements). Similarly, the fact that the measured thicknesses after a long exposure to UV-light (blue, respectively) (25 ± 1 Å, 30 ± 1 Å) are almost equal to the calculated length of the molecule in the *cis* (*trans*, respectively) configuration (~ 26 Å, ~ 31 Å) allow us to estimate that more than 75% of the molecules are in *cis* (*trans*, respectively) isomer after UV (blue, respectively) light exposures. Efficiency of the isomerization in the range 80-99% has been reported.[32] This switching behavior is reproducible and stable. We did not measured more than 10 switching events (due to time limitation, for instance the samples were exposed during 9 hours in total during the 3 cycles of reversible isomeriztion shown in figure 3). However, we do not observe any degradation of the SAM thickness during this exposure to light. This feature is confirmed by CyV showing no change in the surface coverage (2.4x10[14] molecules/cm[2], see above) upon long time exposure to light. This result means that *trans-cis* isomerization occurs without rearrangement at the S-Au anchoring sites as already pointed out by more accurate STM experiments.[33] As in the case of molecules in solution, photoirradiation with 360 nm for 3 h did not lead to significant alteration of the CyV response suggesting i) good stability of the monolayer under irradiation and ii) that *trans* and *cis* isomers show similar electrochemical properties. We have also observed that these SAMs exhibit a long term stability. The switching behavior was still observed after a period of several months at rest in the dark and in a controlled atmosphere (dry nitrogen). Note also that the "*cis*" isomer is stable for a long period (more than 11h) of rest in the dark with no recovering of the "*trans*" isomer (see supporting information). Finally, it is known that photoisomerization efficiency depends on the packing density in the SAM. No switching has been observed for densely-packed SAMs of azobenzene derivatives functionalized with alkythiol chains,[34] while diluted SAMs of azobenzene derivatives in alkylthiol matrix have displayed photoisomerization behaviors.[20, 34] On the other hand, other reports on close-packed SAMs based on single azobenzene derivatives have shown to present *trans-cis* isomerization,[4, 8] and even a collective isomerization of entire molecular-crystalline domains.[33] In that later case, azobenzene derivatives are arranged in a densely-packed (~ 29 Å$^2$ *per* molecule)[33] and crystalline architecture. However their herringbone 2D packing, governed mostly by π-π interactions, enabled the *trans-cis* isomerization. In our case, i) the molecules are less packed (41 Å$^2$ *per* molecule, see above) allowing an easier isomerization; ( ii) the presence of a conjugated bithiophene unit in molecule **1** (AzBT) may also generate π-π interactions that may contributes to collective switching as in ref. [33], however STM imaging at molecular resolution will be further required to confirm this hypothesis.

**Electrical behavior at the nano-scale.**

Current-voltage (I-V) curves were recorded by forming a junction with the metallic tip of a conducting-AFM (see experimental section). Typical I-V curves are presented in figure 4 for a pristine SAM, after irradiation at 360 nm (for 90 min) to switch the molecules to the *cis* isomer, and after exposure to 480 nm light (for 90 min) to switch to the *trans* isomer. From these two later curves, a typical "on"/"off" ratio of ~ 10$^3$ is determined for this typical case (|V|> 0.5 V) irrespective to the bias polarity. For each configuration, we recorded a total of around 40 I-V traces (see experimental section). Histograms of the currents at -1.5 V and 1.5

V are plotted (log-scale) in figures 4b and 4c, respectively. The current distribution follows a log-normal distribution as also observed for alkanethiol SAMs.[35] In the framework of a non-resonant tunnelling transport through the molecular junction, the current is exponentially dependent on the SAM thickness and on the energy barrier height, thus any normal distribution of these parameters leads to a log-normal distribution of the current.[36] This dispersion of the measured current can be due to inhomogeneities in the molecular organization in the SAM and variations of the loading forces (previous experiments on alkyl SAMs show a small increase in the current by a factor ~ 1.7 for a loading force in the range of 20-30 nN).[37] Three peaks (Figs 4b and 4c) are clearly distinguished. For the pristine SAM, two peaks (coming from two sets of measurements on the same SAM) are superimposed and have a log-mean (log-$\mu$) of ~$10^{-9}$ A, and a log-standard deviation (log-$\sigma$) of 1.5. After 90 minutes irradiation at 480 nm (formation of the *trans* isomer), the currents decrease and lead to a peak with log-$\mu$ ~ $1.4 \times 10^{-11}$ A and log-$\sigma$ = 1.9. After 90 minutes irradiation at 360nm (formation of the *cis* isomer), the current increases given log-$\mu$ = $2.1 \times 10^{-8}$ A and log-$\sigma$ = 2.45 (all these values at V = -1.5 V, similar values are obtained at V = 1.5 V, see Fig. 4c). From these histograms, we deduce an "on/off" ratio with log-$\mu$ ~ $1.5 \times 10^3$ and log-$\sigma$ = 4.7. It means that 68% of the devices has an "on/off" ratio between 320 and $7 \times 10^3$.[36] As for the case of contact angle and ellipsometry measurements, we observe intermediate values of the current for the pristine SAM, suggesting that just after its formation, the SAM contains a mixture of the two isomers. These measurements allow us associating the *cis* isomer to the "on" state and *trans* isomer to the "off" state. We repeated these experiments at a macroscopic scale using eutectic GaIn drop contact (see experimental section). We qualitatively reproduced these results (see supporting information). We note that the C-AFM I-V curves are almost symmetric while the molecule itself and the contacts (chemisorbed on one side and mechanical at the tip side) are asymmetric. It has been shown that asymmetries in the molecular junction produce asymmetric I-V.[38-40] However, the chemisorbed Au-S contact is separated from the azobenzene-thiophene moiety by a 4 carbon atoms saturated chain, and we surmize that this short alkyl chain can act as an additional contact resistance that counterbalances the contact asymmetry with respect to the more resistive mechanical molecule-tip side, leading to almost symmetrical I-V curves.

**Electronic structures and first principles calculations.**

To explain the high switching ratio, we analyzed the I-V traces in the *cis* and *trans* isomers with the transition voltage spectroscopy (TVS) method.[41] Recent works show that the TVS method gives a good determination of the position of the molecular orbitals with respect to the Fermi energy of the electrodes.[42, 43] Figures 5a and 5b show typical I-V curves (C-AFM contact) plotted as $\ln(I/V^2)$ vs. $1/V$ for the negative and positive voltages. The bias ($V_T$) at the minimum of $\ln(I/V^2)$ directly determines the energy offset $\Phi = e|V_T|$ between the Fermi energy in the electrodes (the reference of energy being the Fermi energy at the grounded electrode) and the frontier orbitals of the molecules (whether it is the LUMO or HOMO is discussed below). Fig. 5c shows the corresponding histograms for $V_T$ (at both positive and negative biases) well fitted by a normal distribution. At negative bias, we get an energy offset $\Phi_{trans} \approx 1.76$ eV($\pm 0.08$ eV) and $\Phi_{cis} \approx 1.37$ eV ($\pm 0.16$ eV) for the *trans* and *cis* isomers, respectively – these values are the mean values(standard deviation) of the fitted normal distribution shown in figures 5c. At positive bias, we have $\Phi_{trans} \approx 1.91$ eV($\pm 0.04$ eV) and $\Phi_{cis} \approx 1.54$ eV ($\pm 0.18$ eV). The general trend is a reduction of the energy offset for the *cis* isomer. In previous works,[4, 8] the increase in the conductance of the junction for the *cis* isomer was explained by a reduction in the tunnelling barrier length, *i.e.* the length of the molecule. Assuming the same tunnel decay factor β of about 0.45 Å$^{-1}$ for the two isomers in the classical expression of the conductance for a non-resonant tunnelling transport, $G = G_0\exp(-\beta d)$, these authors consistently explained the switching ratio of about 25-30 observed in their experiments (G is the conductance, $G_0$ is an effective contact conductance and d the tunnel barrier thickness, which is the thickness of the SAM). This approach is no longer valid in our case. Explaining a switching ratio > $1.5 \times 10^3$ would require a β value > 1.3 Å$^{-1}$, a too high value for a conjugated molecule (β ≈ 0.4 – 0.6 Å$^{-1}$ for conjugated molecules and 0.8 – 1 Å$^{-1}$ for saturated molecules, i.e. alkyl chains).[44] Moreover, β depends on the tunnelling barrier height, and our observation of a change of Φ definitively excludes the hypothesis of a constant β value. The expression of the tunnelling decay factor β is usually related to barrier height Φ by

$$\beta = \frac{2\sqrt{2m_0 e\Phi}}{\hbar}\alpha$$

where $m_0$ is the electron mass, e the electron charge, $\hbar$ the reduced Planck constant and α an unitless factor accounting for the tunnelling energy barrier shape and/or an effective mass (since we do not know the detail of the potential shape and the effective mass in the junction, $m^* = \alpha^2 m_0$, we take α = 1 which stands for a rectangular barrier and $m^* = m_0$).[45, 46] Assuming the same α value and pre-exponential contact conductance $G_0$ for the two isomers, and taking into account both the changes in the energy barrier and the barrier thickness, the current ratio is given by

$$I_{cis}/I_{trans} = \exp\left(\frac{2\sqrt{2m_0}\alpha}{\hbar}\left(d_{trans}\sqrt{\Phi_{trans}} - d_{cis}\sqrt{\Phi_{cis}}\right)\right) \quad (1)$$

With the data previously measured by ellipsometry ($d_{cis}$ ~ 25 Å; $d_{trans}$ ~ 29 Å) and TVS ($\Phi_{cis}$ ~1.54 eV; $\Phi_{trans}$ ~1.91 eV, see Fig. 4), this ratio is estimated to about $2.1 \times 10^3$. This value is in good agreement with the experimental one; despite it is a crude estimation. The calculated value is highly sensitive to the numerical values in the exponential function, for example if we take $d_{trans}$=30 Å (the error bar is ± 1 Å for the ellipsometry), the ratio increase to about $7.6 \times 10^3$. However, all these estimated numbers are in agreement with the range of the experimental data (68% of the devices has an "on/off" ratio between $3.2 \times 10^2$ and $7 \times 10^3$, see Fig. 4). If $\alpha_{trans} \approx \alpha_{cis}$ seems reasonable, there is no reason having the same $G_0$ for the two isomers. With the molecules in the *trans* isomer, the C-AFM tip is mostly in contact with the terminal $CH_3$ group, while it is mainly in contact with the N=N bonds or the phenyl rings for the *cis* isomer. Since it is known that a saturated bond acts as a tunnel barrier[47-49], the molecules are more weakly coupled electronically with the contact in that case (see DFT calculation below), and it is likely that $G_{0,cis} > G_{0,trans}$. This feature will further increase the *cis/trans* ratio. We believe that this feature can explain the larger ratio than in the previous experiments of Mativetsky and coworkers.[4] In the *cis* conformer, in both experiments the C-AFM tip is mainly in contact with the N=N bond. These authors have measured a current of ~ $10^{-9}$ A (at ± 0.3 V) while we have ~ $7 \times 10^{-9}$ A at the same bias (Fig. 4a). Since the same tip (PtIr with same radius of ~ 20 nm) have been used, our larger value can be due to the higher load force in our experiments. For instance, an increase by a factor ~ 5-10 has been measured between 2 (as in Mativetsky's work) and 20-30 nN (used here, see experimental section) in previous C-AFM experiments.[37] It is difficult to push the comparison further since the molecules used in the two experiments have not exactly the same length, but we can conclude that the conductances $G_{cis}$ are of the same order of magnitude in the two cases. In the *trans* isomer, we measured ~ $10^{-11}$ A (at ±0.3 V), or equivalently ~ $1-2 \times 10^{-12}$ A (renormalized at 2 nN) while Mativetsky et al. reported ~ $7-8 \times 10^{-11}$ A, i.e. a conductance $G_{trans}$ higher by a factor 35-80. Thus, the high on/off ratio in our experiments may be explained by the strong decrease of $G_{trans}$ due to presence, between the phenyl and the tip, of a $CH_3$ end-group acting as a tunnel barrier, while the tip is directly in contact with the phenyl group in the work of Mativetsky and coworkers. In other words, the contact conductance is larger in the *cis* than in the *trans* isomer ($G_{0,cis} > G_{0,trans}$). It is difficult to estimate these values without the help of specific experiments, usually done by measuring the conductance versus the length of a molecule and extrapolating at a null length. In the case of more simple molecules (alkyl chains) a factor 10-100 has been measured between the contact conductance of a chemisorbed contact and a physisorbed one.[50] If we assume $G_{0,cis} \sim G_{0,trans}$ in the experiments of Mativetsky et al.,[4] we can crudely estimate $G_{0,cis}/G_{0,trans} \sim$ 35-80 as discussed above, a factor that can account for a part of the overall 1.5 - 7 x $10^3$ *cis/trans* ratio reported here. Nevertheless, contacts are full parts of the molecular device and we conclude that the high *cis/trans* ratio measured here is an intrinsic phenomenon due to configurationaly-induced changes in the electronic structure of the metal-molecule-metal junction. This can be consistently explained by a combination of the variation of the SAM thickness, change in the electronic structure of the metal-molecules-metal junction, and change in the molecule-electrode coupling.

To probe this later hypothesis further, we performed first principles DFT calculations (see experimental section). Three main DFT results support the higher conductance of the *cis* isomer. i) The frontier orbitals are more delocalized in the *cis* isomer than in the *trans* (figure 6). ii) In the *cis* state, the molecule is more strongly coupled with the electrode as evidenced by the presence of a stronger wavefunction mixing and the higher charge transfer (+0.38 electron on Au) than in the *trans* state (+0.26 e). iii) The energy offset between the Fermi energy and the LUMO is reduced by at least 0.53 eV when the molecule switches from the *trans* to *cis* isomer (figure 6). Although a direct comparison of this last value to the experimental one remains delicate,[51] the trend is consistent with the experimental observation of a reduction of the energy tunnel barrier height. The discrepancy between experiment and theory is more probably related to the fact that a single molecule is used for the calculation and then molecule-molecule interactions are not taken into account. It is known that such interaction leads to a reduction of the HOMO-LUMO gap.[52] Inversely, the energy offset between the Au Fermi energy and HOMO slightly increases when switching from *trans* to *cis* (figure 6).

**Switching kinetics**

Finally, we investigated the switching kinetics by following the time-dependent evolutions of the current (for practical reasons, we measured the kinetics with the eGaIn drop, see experiemental section) under irradiation by blue and UV lights. Before recording the kinetics of the *trans-to-cis (cis-to-trans)* isomerization under illumination with UV (or blue) light, the sample was submitted to lighting for 3 hours with the blue (or UV) light so as to maximize the population of *trans(cis)* isomers, respectively. Each kinetic curve was measured with the same eGaIn tip. The following procedure was repeated: the tip was raised, the surface irradiated for 10 min, and the tip was lowered and put in contact with the surface to measure the I-V characteristic. Since we did not use transparent substrate, the light irradiation and the I-V measurements were not performed simultaneously. From these measurements, we generated the kinetic curves shown in figure 7. Under UV light exposure (*trans-to-cis*), we can distinguish two exponential rate equations with characteristic time constants of 11±1 min and 90±8 min. Under blue light (*cis-to-trans*) only one is observed with a value of 20±1 min. Note that this ratio of about 4.5 is consistent with the difference in light power density (70 µW/cm$^2$ for UV and 250 µW/cm$^2$ for blue lights, respectively). However, the shortest time constant under UV light concerns only a small fraction of the current (i.e. a small number of molecules in the SAM) and, in the following and in the main text, we concentrate on the slowest one which is the limiting factor. We determined the photoionization cross-section by solving the usual first-order rate equations (master equations) given the fraction of molecule in the *cis* (C) and *trans* (T) conformation:

$$\frac{\partial C}{\partial t} = \varphi T \sigma_{TC} - \varphi C \sigma_{CT}$$
$$\frac{\partial T}{\partial t} = \varphi C \sigma_{CT} - \varphi T \sigma_{TC}$$

(2)

with φ the photon flux and $\sigma_{TC}$ and $\sigma_{CT}$ the *trans-to-cis* and *cis-to-trans* photoisomerization cross-sections, respectively. This equation can be solved to give:

- for UV light irradiation with initial boundary conditions: C=0 and T=1

$$C = \frac{\sigma_{TC}}{\sigma_{TC} + \sigma_{CT}} \left(1 - e^{-(\sigma_{TC} + \sigma_{CT})n}\right) \quad (3)$$

- for blue light irradiation with initial boundary conditions: C=1 and T=0

$$T = \frac{\sigma_{CT}}{\sigma_{TC} + \sigma_{CT}} \left(1 - e^{-(\sigma_{TC} + \sigma_{CT})n}\right) \quad (4)$$

where the photon flux and time irradiation are combined to give the photon exposure $n = \varphi\, t$ (total number of photons per unit area). To determine C and T shown in figure 8, we assume that the measured current is roughly proportional to the density of molecules in the *cis* conformation. Thus, we deduced C(t) from data in figure 7a by $C(t) = I(t)/I_\infty$ and T(t) from data in figure 7b by $T(t) = (1 - I(t)/I_0)$ where $I_0$ is the current at t=0 and $I_\infty$ is the saturation current extrapolated from data in figure 7a. C(t) and T(t) versus n are given in figure 8 and fitted with eqs. 3 and 4. Considering (C = 0, T = 1, all molecules *trans*) as the initial conditions of the kinetics equations under UV-light (*trans-to-cis)* and (C = 1, T = 0, all molecules *cis*) under blue-light (*cis-to-trans*) are strong hypothesizes. While we cannot exclude that a fraction of molecules in the SAMs cannot switch, nor that the *cis-to-trans* and *trans-to-cis* transformation are uncomplete, these initial conditions seem reasonable according to thickness measurements by ellipsometry (see above, section 1). Table 1 gives the fitted (fits are the full lines in Fig. 8) values of $\sigma_{TC}$ and $\sigma_{CT}$. They are compared with the values of two others experiments on monolayers of azobenzene derivatives.[12, 53, 54] Our values are very close to those measured by sum-frequency generation vibrational spectroscopy on a SAM of azobenzene linked to Au surface through a linker system composed of a tripodal unit and an adamantane core.[54] Compared to the data in Ref. [53], we obtain larger photoionization cross-sections and a clear asymmetrical behavior, $\sigma_{TC} > \sigma_{CT}$ under UV light and $\sigma_{CT} > \sigma_{TC}$ under blue light, while $\sigma_{CT} \approx \sigma_{TC}$ in Ref. [53]. This feature may be due to differences in the molecular arrangement in the monolayers in both experiments. In ref. [53], the azobenzenes are lying flat on the surface and are decoupled by short ($\approx$ 3.2 Å) tert-butyl legs. On the contrary, in our case, as in Ref. [54], the photoisomerizable azobenzene is strongly "mechanically" and "electronically" decoupled from the metal surface by a long linker ($\approx$ 14 Å for the 4 carbon atoms alkyl and the bithiophene), and the measured photoionization cross-sections are on a par with the known value for azobenzene family in solution. Such results illustrate the important role of the spacer in the dynamical behavior of azobenzene derivatives self-assembled on a solid surface.

**Conclusions**

In conclusion, self-assembled monolayers of a new azobenzene-thiophene derivative exhibit high "on/off" conductance ratio upon configurational change, with the higher conductance state associated to a *cis* isomer of the azobenzene moiety. In the framework of a non-resonant tunneling transport through the metal-molecule-metal (at low bias) this ratio is satisfactorily explained by considering a combined modification of both the tunnel barrier length (i.e. decrease in the length of the molecule in the *cis* form) and a lowering of the energy position of the LUMO with respect of the electrode Fermi energy for the *cis* isomer. First principles DFT calculations support our conclusions, and clearly demonstrate a better delocalization of frontier orbitals, as well as a stronger electronic coupling between the azobenzene moiety and the electrodes, in the *cis* configuration than in the *trans* one. These results allow us to envision these molecules as building blocks for fast (with $\sigma \sim 10^{-18}$ cm$^2$ switching time ~ 1-10 µs could be achieved, in principle, with a high intensity light source) solid-state molecular switch and memory devices with a high on/off conductance ratio.

**Experimental Section**

*Synthesis of the self-assembled monolayers*: The synthesis of titled compound **1** (AzBT) is obtained in 80% yields from precursor **2** after a saponification reaction of the thioester group by treatment with cesium hydroxide followed by addition of hydrochloric acid (Scheme 1). Details on the synthesis and characterization of these molecules will be given elsewhere.[55]

We prepared gold Au(111) substrates by evaporating 10 nm of titanium to promote adhesion and 100 nm of gold onto cleaned silicon substrate using an e-beam evaporator. We used a low deposition rate (1 Å.s$^{-1}$) at 10$^{-8}$ torr to minimize the roughness. Before grafting the molecules, we cleaned the Au substrates by piranha attack (1/3 H$_2$O$_2$ 30% and 2/3 concentrated H$_2$SO$_4$ 96%) during 15 minutes, followed by an immersion in a dilute aqua regia solution (3:1:16 HCl 37%:HNO$_3$ 65%:H$_2$O) during 5 minutes to reduce gold roughness (≈ 1.7 nm rms). For the SAM fabrication, we exposed this freshly cleaned gold surface to 1 mM solution of **1** in dichloromethane during 72 h. The reaction took place in a glove box with a controlled atmosphere (N$_2$ with < 1ppm of water and oxygen) and in the dark. Then, we rinsed the treated substrates with dichloromethane, followed by a cleaning in an ultrasonic bath of dichloromethane during 5 minutes. A stable cyclic voltammetry (CyV) trace of monolayer of **1** exhibits a reversible one-electron oxidation wave at $E_{pa}^1$ = 0.98 V associated with a reduction peak potential at 0.95 V measured during the reduction back sweep. CyV analysis of the monolayer has been carried out at different scan rates ranging from 50 to 8000 mV/s. The linear variation of the peak current versus scan rate and the invariance of $E_{pa}^1$ = 0.98 V with scan rate confirms that molecules **1** are immobilized on the electrode surface. The surface coverage has been determined by integration of the voltammetric peak after correction for double layer charge. A value of ca. $4 \times 10^{-10}$ mol.cm$^{-2}$ was obtained for monolayer of **1**, indicative of a densely close-packed monolayer.

*Contact-angle measurements.* We measured the water contact angle with a remote-computer controlled goniometer system (DIGIDROP by GBX, France). We deposited a drop (10-30 µL) of desionized water (18MΩ.cm$^{-1}$) on the surface and the projected image was acquired and stored by the computer. Contact angles were extracted by contrast contour image analysis software. These angles were determined few seconds after application of the drop. These measurements were carried out in a clean room (ISO 6) where the relative humidity (50%) and the temperature (22°C) are controlled. The precision with these measurements are ± 2°.

*Spectroscopic ellipsometry.* We recorded spectroscopic ellipsometry data in the visible range using an UVISEL (Jobin Yvon Horiba) Spectroscopic Ellipsometer equipped with a DeltaPsi 2 data analysis software. The system acquired a spectrum ranging from 2 to 4.5 eV (corresponding to 300 to 750 nm) with intervals of 0.1 eV (or 15 nm). Data were taken at an angle of incidence of 70°, and the compensator was set at 45.0°. We fitted the data by a regression analysis to a film-on-substrate model as described by their thickness and their complex refractive indexes. First, we recorded a background before monolayer deposition for the gold coated substrate. Secondly, after the monolayer deposition, we used a 2 layers model (substrate/SAM) to fit the measured data and to determine the SAM thickness. We used the previously measured optical properties of the gold coated substrate (background), and we fixed the refractive index of the organic monolayer at 1.50. The usual values in the literature for the refractive index of organic monolayers are in the range 1.45-1.50.[56, 57] We can notice that a change from 1.50 to 1.55 would result in less than 1 Å error for a thickness less than 30 Å. We estimated the accuracy of the SAM thickness measurements at ± 1 Å.

*XPS measurements.* We performed XPS measurements to control the chemical composition of the SAMs and to detect any contaminant. We used a Physical Electronics 5600 spectrometer fitted in an UHV chamber with a residual pressure of 2x10$^{-10}$ Torr. High resolution spectra were recorded with a monochromatic AlKα X-ray source (hυ=1486.6 eV), a detection angle of 45° as referenced to the sample surface, an analyzer entrance slit width of 400mm and with an analyzer pass energy of 12 eV. In these conditions, the overall resolution as measured from the full-width half-maximum (FWHM) of the Ag $3d_{5/2}$ line is 0.55 eV. Semi-quantitative analysis were completed after standard background substraction according to Shirley's method.[58] Peaks were decomposed by using Voigt functions and a least-square minimization procedure and by keeping constant the Gaussian and Lorentzian broadenings for each component of a given peak.

*Electrical measurements.* We performed current-voltage measurements by conducting-Atomic Force Microscopy (C-AFM) in ambient air (Dimension 3100, Veeco), using a PtIr coated tip (tip radius of curvature less than 25nm, force constant in the range 0.17-0.2N/m). Placing the conducting tips at a stationary point contact formed nano-junctions. A square grids of 10x10 is defined with a lateral step of 2 nm. At each point, 10 I-V curves are acquired and averaged. Out of the 100 I-V traces, some are eliminated due to lateral drift of the C-AFM set-up and about 40 I-V "mechanically stable" traces are used for statistics. The load force was adjusted in the range 20-30 nN and measured by force-distance curves with the controlling software of the Dimension 3100. The bias was applied on the Au substrate and the tip was grounded through the input of the current amplifier. We focused the light from a xenon lamp to an optical fiber. We used two dichroic filters centred at 360 nm and 480 nm. At the output of the optical filter, the SAMs were irradiated on about 1 cm² at power density of 70 $\mu$W/cm² (at 360 nm) and 250 $\mu$W/cm² (at 480nm). Due to the size and design of the Dimension 3100, it is not easily possible to bring the optical fiber close to the C-AFM tip, thus the light irradiation was not done with the C-AFM tip contacting the monolayer. We moved the sample mounted on the chuck back and forth between AFM tip area and the illumination area located nearby on the Dimension 3100 platform. We also used Eutectic GaIn drop contact (eGaIn 99.99%, Ga:In; 75.5:24.5 wt% from Alfa Aesar). We used a method close to the one developed by Chiechi et al.[59]. We formed a drop of eGaIn at the extremity of a needle fixed on a micromanipulator. By displacing the needle, we brought the drop into contact with a sacrificial surface, and we retracted the needle slowly. By this technique, we formed a conical tip of eGaIn with diameter ranging from 50 to 200 $\mu$m (corresponding to contact area ranging from $10^{-5}$ to $10^{-3}$ cm$^2$). This conical tip was then put into contact with SAM (under control with a digital video camera). Voltage was applied on the eGaIn drop. The contact area was also determined by measuring the 1MHz capacity of the junction and assuming a relative permittivity of $\varepsilon_R=2.2$ for the SAM.

*First principles calculations.* Quantum chemical density functional theory (DFT) calculations were performed with the NWChem software package.[60] We used a 6-31G basis set for carbon, nitrogen, sulphur and hydrogen atoms in conjunction with the B3LYP functional for exchange and correlation. For gold atom, the LANL2DZ basis set was used to describe the 19 valence and 60 core electrons. The isolated molecular structure of the AzBT was fully optimized without symmetry constraints using the quasi-Newton method until a gradient convergence factor better than $10^{-5}$ Hartree/Bohr was reached. Several starting geometries were considered during this optimization process, and only the most stable final structures (*cis* and *trans*) were conserved. The electronic structures of the Au$_{11}$-AzBT ($^2A_1$ states) were obtained from the optimized AzBT molecule where we placed a fixed Au$_{11}$ cluster at virtually the same distance (0.2 nm) from *cis* or *trans* AzBT molecule. The relative stability of the species was calculated with respect to the ground state geometry.

**Acknowledgements.** This work was financially supported by ANR-PNANO under the OPTOSAM project (number ANR-06-NANO-016).

**Supporting information available.** Current-voltage curves at a macroscopic scale (Hg drop) and corresponding discussion. Figures S1 and S2. This material is available free of charge via the Internet at http://pubs.acs.org.

**References.**

36. If X is a random variable with a normal distribution, then Y = exp(X) has a log-normal distribution; likewise, if Y is log-normally distributed, then log(Y) is normally distributed. If m and s are the mean value and standard deviation of normal distribution of log(Y), we denote log-µ the log-mean (=$10^m$) and log-σ the log standard deviation (=$10^s$) of the log-normal distribution of Y. All results are given as log-µ(log-σ) otherwise specified. Thus, 68% of the measured data are between (log-µ)/(log-σ) and (log-µ)x(log-σ).

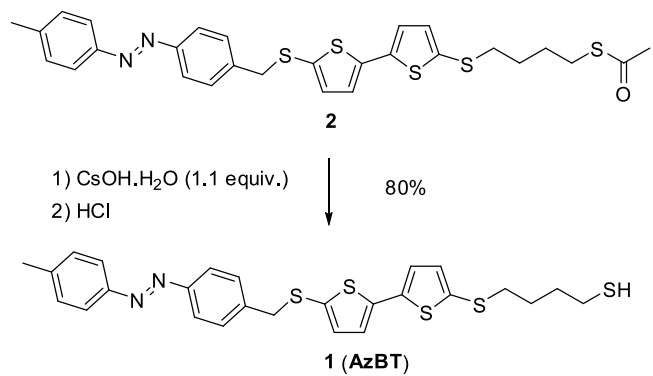

**Scheme 1.** Chemical structures of the molecules

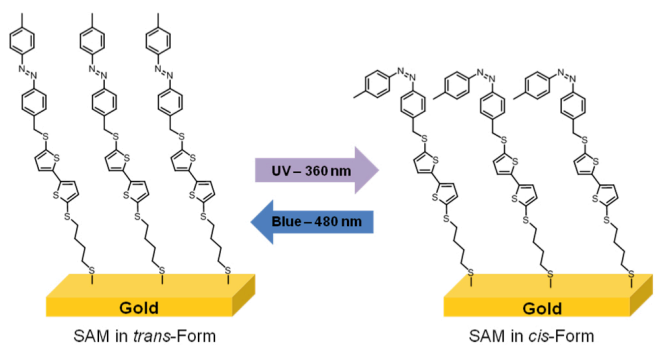

**Figure 1.** Schematic view of the SAM with the AzBT molecules in the *trans* and *cis* isomers.

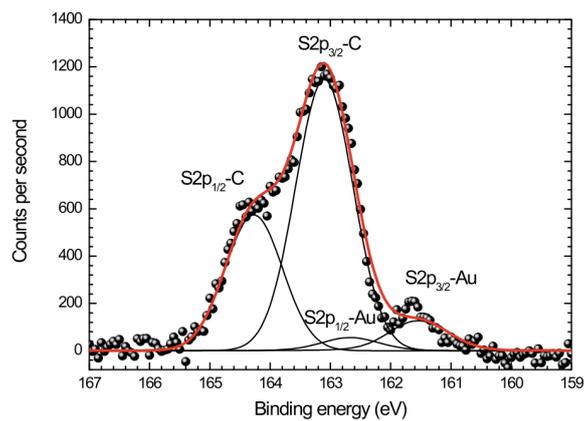

**Figure 2.** Curve-fitted high-resolution XPS spectra for the $S_{2p}$ region of SAM grafted on gold surface. The apparition of two doublets is due to the presence of sulfur atoms linked with gold atoms or linked with carbon atoms, and each doublet results from the spin–orbit splitting of the $S_{2p}$ level.

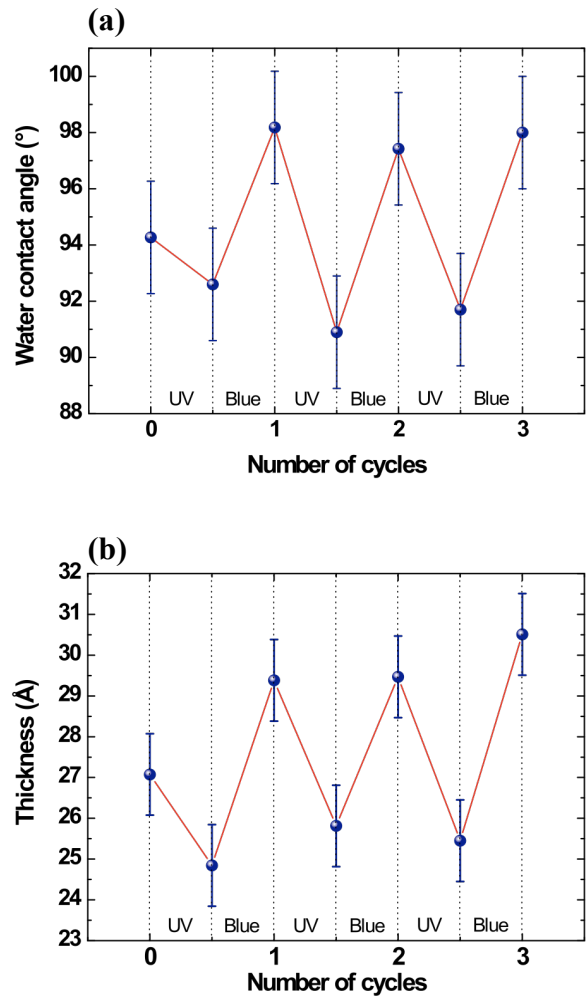

**Figure 3.** **(a)** Evolution of the water contact angle of the SAM and **(b)** evolution of the thickness of the SAM measured by Spectroscopic ellipsometry as a function of the irradiation during 3 cycles. After formation, the SAM was alternatively exposed to UV light (360nm) and visible light (480nm) during 90 minutes.

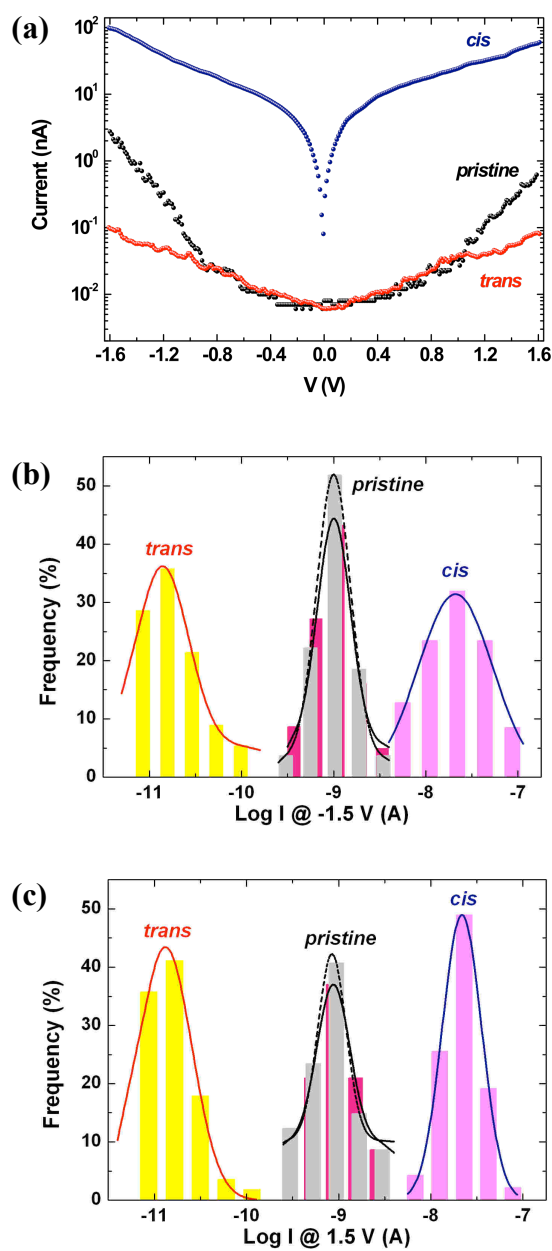

**Figure 4. (a)** Typical current (abs. value)-voltage (I-V) curves measured by C-AFM (load force 20-30 nN, voltage applied on the Au substrate of the sample) on a pristine SAM immediately after its formation, after irradiation at 360 nm during 90min to switch the molecules in their *cis* isomer and after irradiation at 480 nm (90min) to switch in the *trans* isomer. The curves are averaged over 10 I-V traces taken at the same C-AFM tip location. **(b)** Histograms of the current (in a log scale) for the same experiments (~ 40 I-V traces for each peaks). Three distinguishable log-normal distribution of the current are observed with log-µ(log-σ) values of ~ $10^{-9}$(1.5) A, $1.4\times10^{-11}$(1.9) A and $2.1\times10^{-8}$(2.45) A for the pristine SAM, *trans* and *cis* forms, respectively. **(c)** Same at 1.5 V with log-µ(log-σ) values of ~ $8.8\times10^{-8}$(1.5) A, $1.3\times10^{-11}$(1.9) A and $2.2\times10^{-8}$(1.6) A for the pristine SAM, *trans* and *cis* forms, respectively.

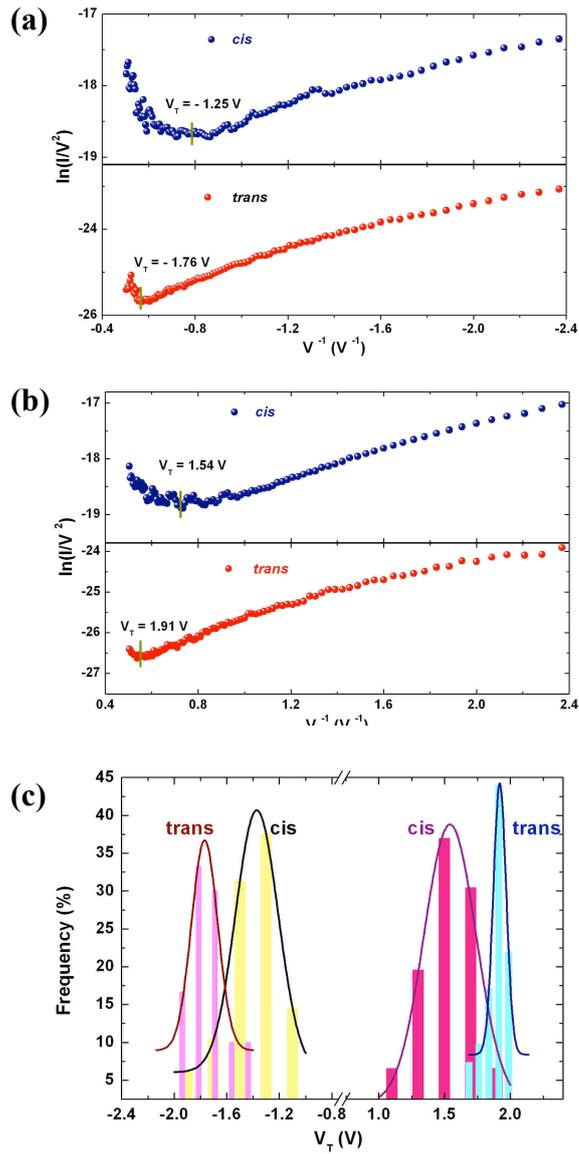

**Figure 5.** Plots of ln(I/V²) versus 1/V for the *cis* and *trans* configurations at **(a)** negative and **(b)** positive voltages. **(c)** Histograms of the transition voltage $V_T$ (voltage at the minimum of the ln(I/V²) which determines the energy offset $\Phi = e\,|V_T|$. For C-AFM measurements, at negative voltages, the normal distribution (lines) gives $\Phi_{trans}$ = 1.76 eV (± 0.08 eV) – mean value μ with the standard deviation σ in brackets – for the *trans* isomer, and $\Phi_{cis}$ = 1.37 eV (± 0.16 eV) for the *cis* isomer. At positive bias, we get: $\Phi_{trans}$ =1.91 eV (± 0.04 eV) and $\Phi_{cis}$ = 1.54 eV (± 0.18 eV).

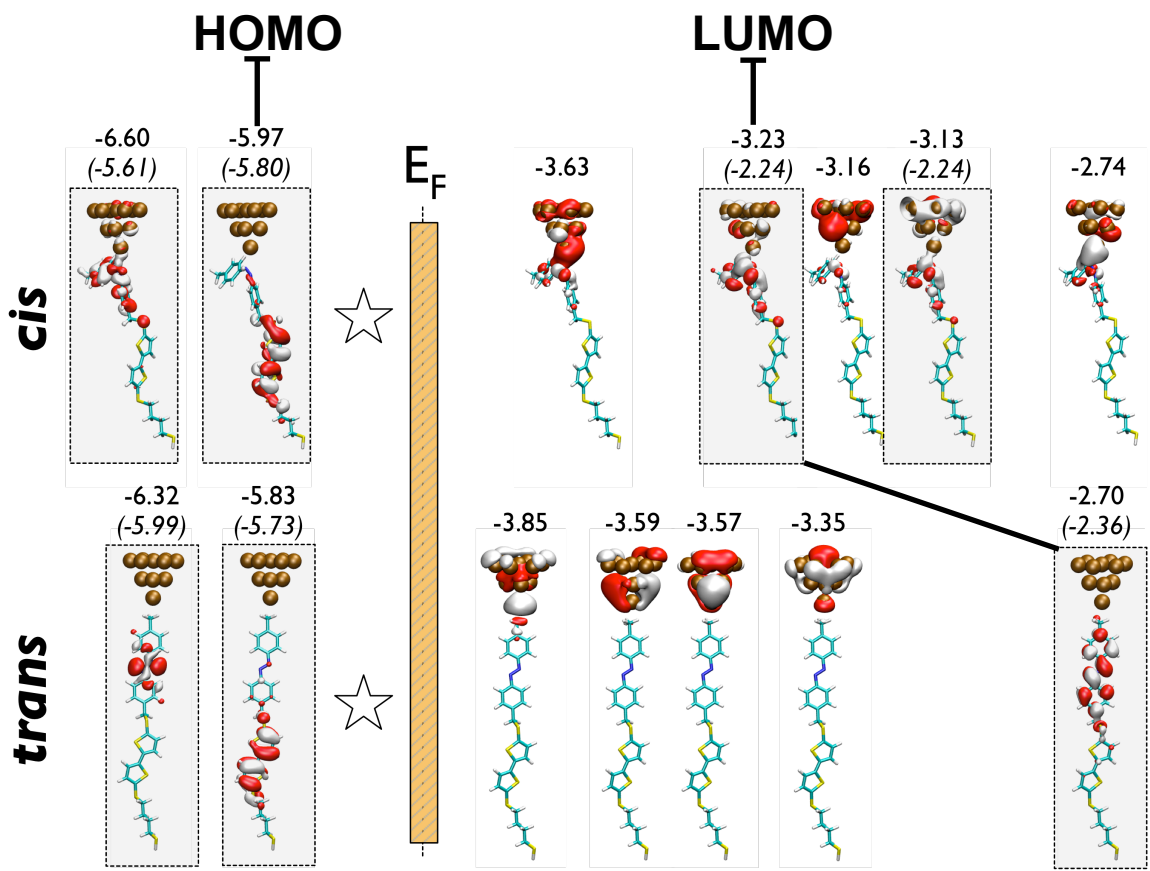

**Figure 6.** DFT calculations. Localization of the main frontier molecular orbitals (MOs) for the *trans* and *cis* isomers contacted by a gold tip as obtained by DFT calculations. We identified the MOs that originate from frontier MOs of isolated molecule by a light gray box. The star symbol indicates that two MOs centered on the Au11 electrode are not shown here. The DFT energy levels are given (in eV) for the comple Au/molecule while values in parenthesis are the calculated levels for the isolated molecule.

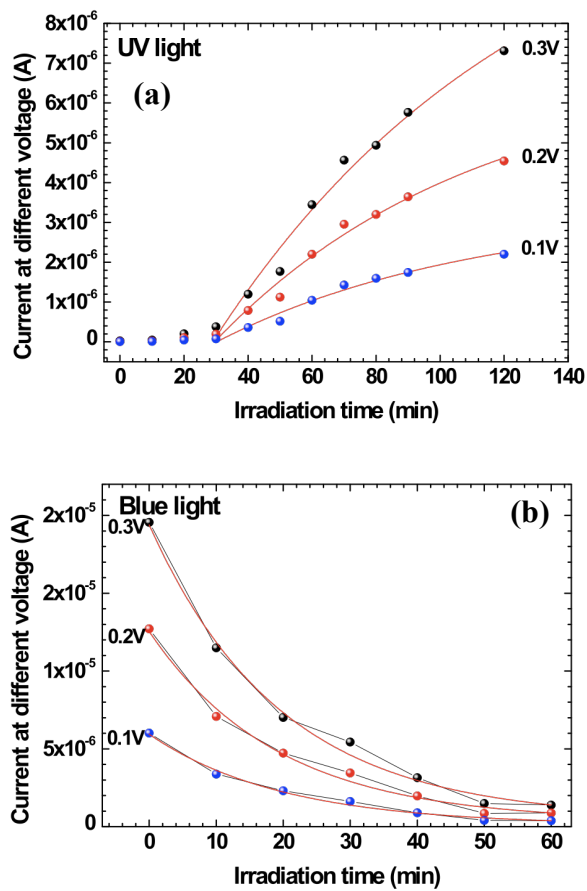

**Figure 7.** Evolution versus time of the current measured at different voltages (0.3V; 0.2V and 0.1V) under light exposition: **(a)** UV: *trans-to-cis* isomerization and **(b)** blue light: *cis-to-trans* isomerization. The lines correspond to the fits with a first-order rate equation (see text). Under, the UV light, the first 30 min are not taken into account (see text).

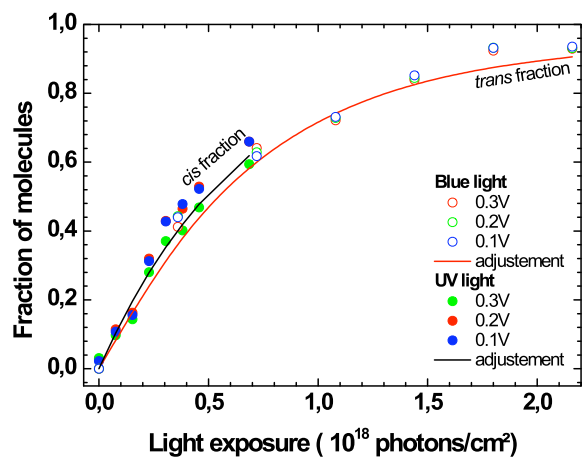

**Figure 8.** Evolution of the population of molecules in the *cis* and *trans* isomers as a function of the flux of photons irradiating the samples under a UV-light (360 nm) and a blue light (480 nm), respectively. Before recording the *trans to cis* kinetics (at 360 nm), we illuminated the sample at 480 nm during 3 hours to maximize the population of molecules in the *trans* isomer, and inversely before recording the *cis to trans* kinetics. The lines are the best fits of equations 3 and 4 with the photoionization cross-sections given in table 1.

**Table 1.** Values of the *trans-to-cis* $\sigma_{TC}$ and *cis-to-trans* $\sigma_{CT}$ photoionization cross-sections. In our experiments, uncertainties represent the dispersion for measurements taken at various voltages applied on the samples.

|  | $\sigma_{TC}$ (cm²) | $\sigma_{CT}$ (cm²) | Refs |
|---|---|---|---|
|  | $(1.5\pm0.2)\times10^{-18}$ | $(5.4\pm2.0)\times10^{-20}$ | This work |
| UV light (360nm) | $2.3\times10^{-23}$ | $2.3\times10^{-23}$ | Ref. 53 |
|  | $(4\pm1)\times10^{-18}$ | / | Ref. 54 |
|  | $(3.9\pm0.8)\times10^{-20}$ | $(1.4\pm0.2)\times10^{-18}$ | This work |
| Blue light (480nm) | $1.7\times10^{-23}$ | $2.3\times10^{-23}$ | Ref. 53 |
|  | / | $(2.5\pm0.9)\times10^{-19}$ | Ref. 54 |

**Graphical ToC:**

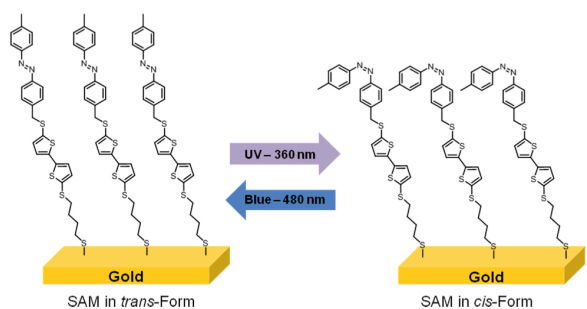

# High on-off conductance switching ratio in optically-driven self-assembled conjugated molecular systems

*Kacem Smaali,[1] Stéphane Lenfant,[1] Sandrine Karpe,[2] Maïténa Oçafrain,[2] Philippe Blanchard,[2] Dominique Deresmes,[1] Sylvie Godey,[1] Alain Rochefort,[3] Jean Roncali[2] and Dominique Vuillaume[1]*

1) Institute for Electronics Microelectronics and Nanotechnology, CNRS & University of Lille, Molecular Nanostructures & Devices Group, B.P. 60069, 59652, Villeneuve d'Ascq , France.

2) Linear Conjugated Systems group, CIMA, CNRS & University of Angers, 2 Bd Lavoisier, 49045, Angers, France.

3) Département de génie physique, École Polytechnique de Montréal, Montréal, Canada H3C 3A7.

* Corresponding authors: dominique.vuillaume@iemn.univ-lille1.fr; stephane.lenfant@iemn.univ-lille1.fr

## *Supporting information*

### Current-voltage at a macroscopic scale.

I-V curves were also recorded by forming junctions with a drop of a mouldable eutectic Gallium Indium (eGaIn, see experimental section, main text). Figure S1-a shows typical I-V curves. Figures S1-b and S1-c show the histograms for the current in the *cis* and *trans* states and of the *cis/trans* current ratios at several biases. As for the case of C-AFM measurements, the currents are log-normal distributed. A typical ratio of log-$\mu$(log-$\sigma$) =180(1.3) is obtained at – 0.6 V, while the highest value of 750(1.16) is measured at – 1 V. For positive voltages, the *cis/trans* ratio is almost constant with a slight maximum of 110(1.3) at 0.6 V. The I-V characteristics measured by this technique are highly reproducible for several measurements of the same junction, as well as between different junctions obtained by moving the eGaIn at various places on the SAMs. The asymmetry for the I-V curves and for the "on/off"ratio with the eGaIn contact may be due to the presence of a very thin (0.5 - 1 nm) layer of $Ga_2O_3$ semiconducting oxide on the surface of the eGaIn drop.

The *cis/trans* current ratio is lower for the macroscopic eGaIn contact than for C-AFM. A similar trend was also observed for a slightly different SAM of azobenzene derivatives, but with a low switching ratios (~ 30 by C-AFM and ~ 25 by Hg drop).[1] In agreement with these authors, it is likely that our smaller ratio for the macroscopic measurements is due to the fact that the measurement includes the contribution of some "dead zone" (non switching domains) of the SAMs and "defects" (such as grain boundaries between switching domains). These features tend to decrease the switching ratio. On the contrary, the typical size of the contact using C-AFM is of the same order of magnitude as the switching domain (few tens of nanometers)[2] and the maximum switching dynamic is obtained. This hypothesis is also consistent with the larger dispersion observed for the "on/off" ratio with C-AFM measurements (log-$\sigma$=4.7) than with the eGaIn contact (log-$\sigma$=1.3) because C-AFM can directly probe the inhomogeneities in the SAMs while they are averaged with the eGaIn contact. We also note that the *cis* isomer is stable in dark and at room temperature for more than 10 hours without experiencing any thermal *cis-to-trans* isomerisation. Figure S2 shows a cycle of reversible *cis/trans* isomerization including a long period of rest in the *cis* configuration. This experiment demonstrate that the *cis* state is stable over a long period of time (11.5 hours) in dark, at room temperature and under applied bias (0.5 V in Fig. S2). In our case, the *cis* isomer does not experience a slow, thermally activated, back isomerization to the more stable *trans* conformation (by about 0.7 eV from DFT). This feature implies that the AzBT molecule can be used as light-driven molecular memory.

Figures S3-a and S3-b shows the TVS plots with the eGaIn drop contact. Fig. S3-c show the corresponding histograms for $V_T$ (at both positive and negative biases) well fitted by a normal distribution. At negative bias, we get $\Phi_{trans} \approx 0.49$ eV($\pm 0.07$ eV) and $\Phi_{cis} \approx 0.24$ eV ($\pm 0.01$ eV) for the *trans* and *cis* isomers, respectively. At positive bias, we observe a smaller barrier reduction from $\Phi_{trans} \approx 0.51$ eV (0.06 eV) to $\Phi_{cis} \approx 0.47$ eV (0.01 eV). As for the C-AFM measurements, the general trend is a reduction of the energy offset for the *cis* isomer. The fact that the $V_T$ values for the C-AFM are systematically larger than for the eGaIn measurements (an average difference between 1.1 and 1.4 eV) is in agreement with the smaller work function (of about 0.9 eV) in this latter case.

M.; Samori, P.; Mayor, M.; Rampi, M. A., Light-powered electrical sitch based on cargo-lifting azobenzene monolayers. *Angew. Chem. Int. Ed. Engl.* **2008,** *47*, 3407-3409.
2. Pace, G.; Ferri, V.; Grave, C.; Elbing, M.; von Hänisch, C.; Zharnikov, M.; Mayor, M.; Rampi, M. A.; Samori, P., Cooperative light-induced molecular movements of highly ordered azobenzene self-assembled monolayers. *Proc. Natl. Acad. Sci. USA* **2007,** *104* (24), 9937-9942.

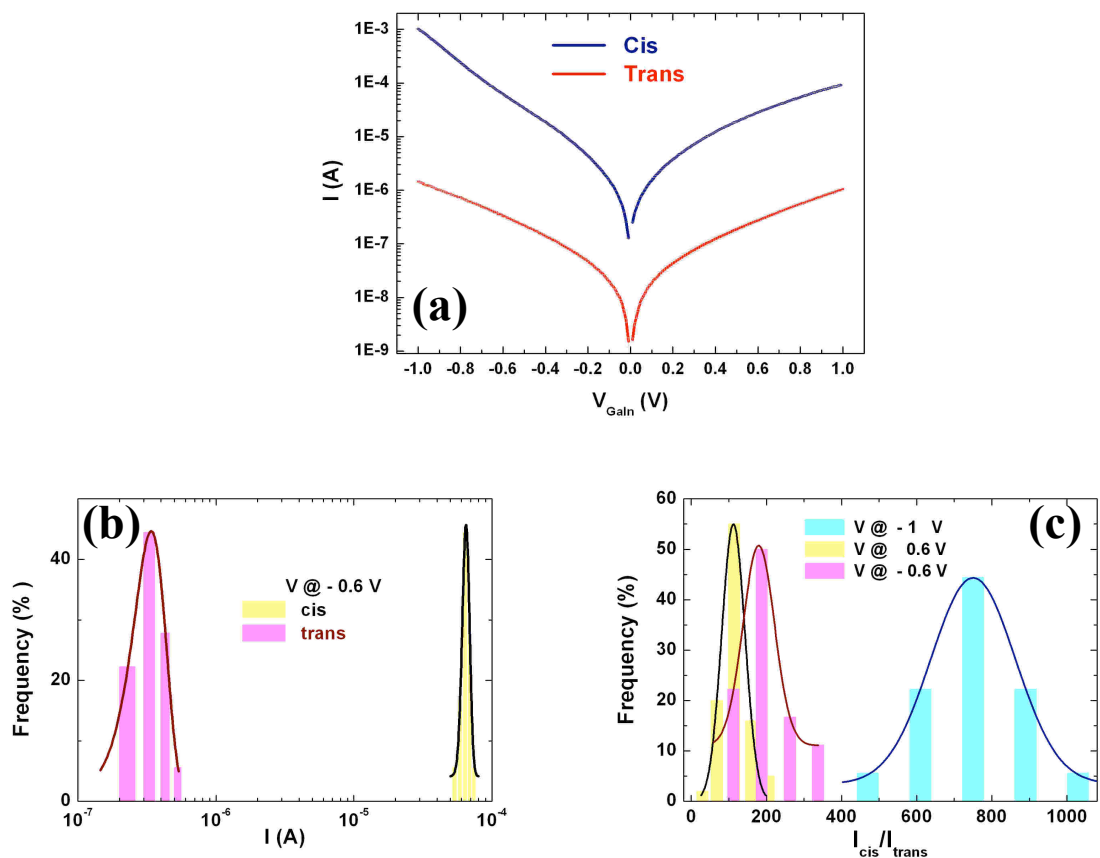

**Figure S1. I-V for junctions with eGaIn drop contact**. **a)** Typical I-V curves for the *cis* and *trans* isomers (V applied on eGaIn drop). **(b)** Histograms of the current (at -0.6 V) for the two isomers. The current is log-normal distributed and we have $3.4 \times 10^{-7}$ (1.35) A for the *trans* form – these values are log-mean (log-standard deviation) – and $6 \times 10^{-5}$ (1.07) A for the *cis* isomer. **(c)** Histograms for the *cis/trans* current ratio at several voltages. The highest ratio is observed for negative bias with $7.5 \times 10^{2}$ (1.16) at – 1V, $1.8 \times 10^{2}$ (1.3) at – 0.6 V and $1.1 \times 10^{2}$ (1.3) at 0.6 V.

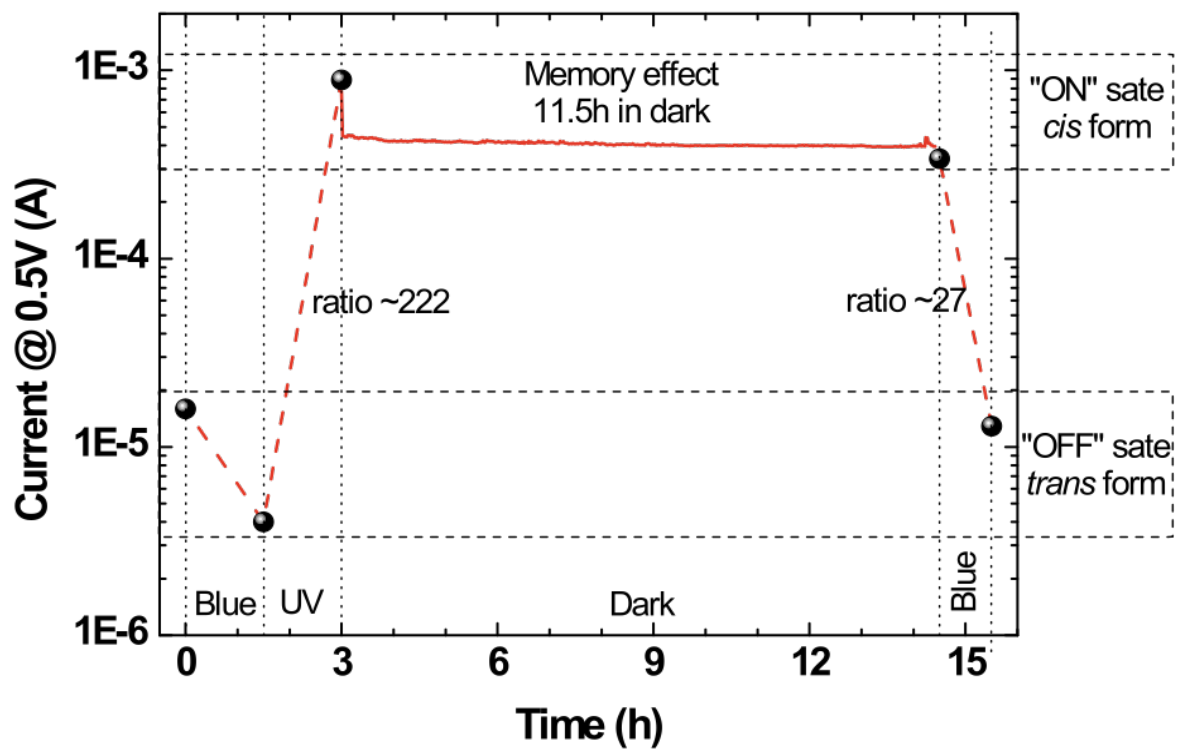

*Figure S2.* Current measured at 1V in function of time under various light expositions successively: blue 90min, UV 90min, dark 11.5h and blue 60min. During the dark period, the current was followed in function of the time at 0.5V.

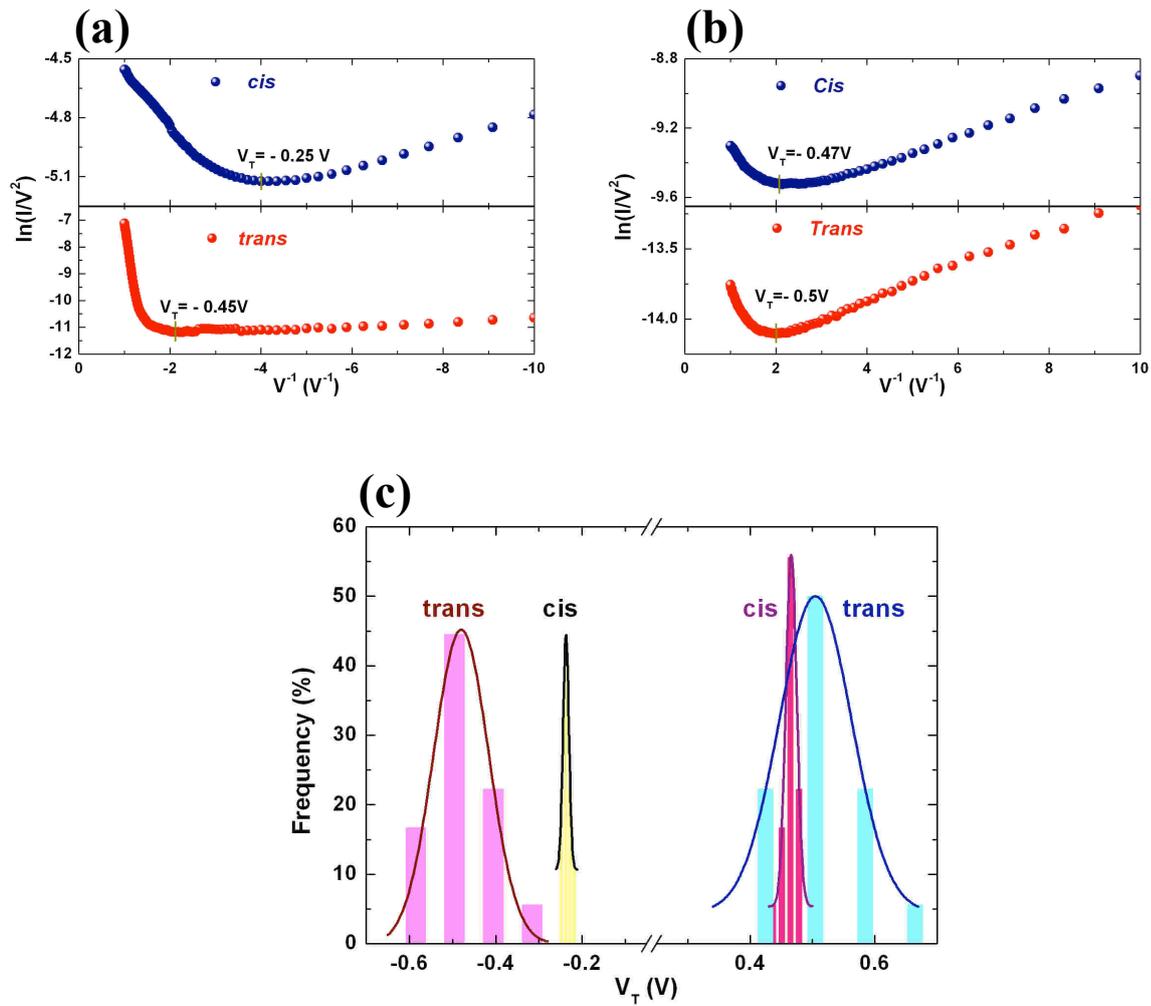

**Figure S3. Transition voltage spectroscopy.** Plots of ln(I/V²) versus 1/V for the *cis* and *trans* configurations at **(a)** negative and **(b)** positive voltages for eGaIn measurements. **c)** Histograms of the transition voltage $V_T$ (voltage at the minimum of the ln(I/V²) which determines the energy offset $\Phi = e\,|V_T|$. At negative voltages, $\Phi_{trans}$ = 0.49 eV (±0.07 eV) and $\Phi_{cis}$ = 0.24 eV (±0.01 eV), and at positive bias: $\Phi_{trans}$ = 0.51 eV (±0.06 eV) and $\Phi_{cis}$ = 0.47 eV (±0.01 eV).

5